\documentstyle[epsfig]{article}
\newcommand{\be}{\begin{equation}}
\newcommand{\ee}{\end{equation}}

\begin{document}
\title{Semiclassical Calculation of Diffractive Parton Densities}

\author{A. Hebecker\\ {\it
D.A.M.T.P., Cambridge University, Cambridge CB3 9EW, England}}

\maketitle

\begin{abstract}
In this talk the relation between the semiclassical approach and the concept 
of diffractive parton densities is discussed. The proton rest frame 
calculation is organized in a way that exhibits the hard partonic cross 
section and the diffractive parton density as the two fundamental 
ingredients. The latter one is a non-perturbative quantity which, in the 
present model, is explicitly given by integrals of non-Abelian eikonal 
factors in the colour background field. 
\end{abstract}

\section*{Introduction}
The presence of a large diffractive component of the small-$x$ 
deep-inelastic scattering structure functions \cite{ex} suggests an 
interesting interplay of soft and hard physics. 

Consider the process of diffractive deep-inelastic scattering in the rest 
frame of the proton. In this frame a partonic fluctuation of the photon with 
virtuality $Q^2$ scatters off the proton, producing a final state with 
invariant mass $M$. In the case of a large transverse size of the 
fluctuation, corresponding at leading order to aligned jet configurations 
\cite{bk}, a leading twist diffractive cross section arises. This can be 
seen explicitly from the semiclassical calculation, where the target proton 
is modelled by a classical colour field \cite{wb}. 

Hard diffraction can also be approached along the lines of inclusive 
deep-inelastic scattering, working in a frame where the incoming proton is 
fast. Focussing strictly on the leading power of $Q$, diffractive processes 
are then expected to allow a parametrization within the partonic approach, 
with diffractive parton distributions being the fundamental non-perturbative 
objects \cite{is}. 

In the following the leading order derivation of the above partonic picture 
in the rest frame of the proton shall be outlined. More specifically, it 
will be shown how the semiclassical calculation can be factorized into hard 
and soft part and how a model for diffractive parton densities naturally 
arises from the soft part of the semiclassical calculation \cite{ah}.

\section*{Outline of the calculation in the scalar case}
\begin{figure}[ht]
\centerline{\epsfig{file=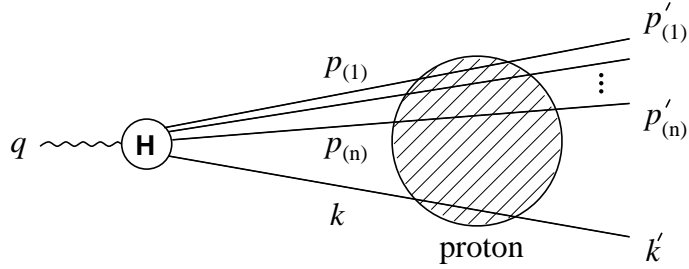,width=9cm}}
\vspace{10pt}
\caption{Hard diffractive process in the proton rest frame. The soft parton 
with momentum $k$ is responsible for the leading twist behaviour of the 
cross section. }
\label{process}
\end{figure}

The essential ideas of the calculation are most easily understood in a model 
containing scalar coloured particles. Consider the fluctuation of the 
incoming photon into a set of scalar partons which interact independently 
with the proton colour field (see Fig.~\ref{process}).

Assume furthermore that the transverse momenta $p_{(j)\perp}' (j=1...n)$ are 
hard, i.e. ${\cal O}(Q)$. A leading twist contribution to diffraction can 
arise only if the transverse momentum of the remaining parton is small,
$k_\perp'\sim \Lambda_{\mbox{\scriptsize QCD}}$ \cite{wb}. 

The standard cross section formula for the scattering off a static external 
field reads
\be
d\sigma=\frac{1}{2q_0}\,|T|^2\,2\pi\delta(q_0-q_0')\,dX^{(n+1)}
\,,\quad\mbox{where}\quad q'=k'+\mbox{$\sum$}p_{(j)}'\, .
\ee
All momenta are given in the proton rest frame, $T$ is the amplitude 
corresponding to Fig.~\ref{process}, and $dX^{(n+1)}$ is the usual phase 
space element for $n+1$ particles.

In the high energy limit each of the particles scatters off the field with 
an effective vertex proportional to $p_0\delta(p_0'-p_0)\tilde{U}(p_\perp'-
p_\perp)$, where $\tilde{U}$ is the Fourier transform of the 
non-Abelian eikonal factor $U(x_\perp)$. This eikonal factor describes the 
colour rotation experienced by a parton penetrating the external field at 
transverse position $x_\perp$.

The resulting amplitude is given by
\[
i\,2\pi\delta(q_0-q_0')\,T\!=\!\int\!T_H\,\prod_j\!\left(\frac{i}{p_{(j)}^2}
\,2\pi\delta(p_{(j)0}'-p_{(j)0})\,2p_{(j)0}\,\tilde{U}(p_{(j)\perp}'-
p_{(j)\perp})\,\frac{d^4p_{(j)}}{(2\pi)^4}\right) 
\]
\be
\hspace*{5cm}\times\left(\frac{i}{k^2}\,2\pi\delta(k_0'-k_0)\,2k_0\,
\tilde{U}(k_\perp'-k_\perp)\right)\,,
\ee
where $T_H$ stands for the hard part of the diagram in Fig.~\ref{process}.

The integrations over the light-cone components $p_{(j)+}$ can be performed 
using the energy $\delta$-functions. The $p_{(j)-}$-integrations are 
performed by picking up the poles of the propagators $1/p_{(j)}^2$. Since 
the external field is smooth and $T_H$ is dominated by the hard scale, $n-1$ 
of the $n$ transverse momentum integrations can be performed trivially. This 
is not the case for the last integration which will necessarily be sensitive 
to the small off-shellness $k^2$. However, the $n-1$ performed integrations 
ensure that the eikonal factors associated with the high-$p_\perp$ partons 
are evaluated at the same transverse position. The resulting colour 
structure of the amplitude, after projection on a colour singlet final 
state, involves the trace of
\be
W_{x_\perp}(y_\perp)=\left(U(x_\perp)^\dagger U(x_\perp+y_\perp)-1
\right)\,.
\ee
It is intuitively clear that only two eikonal factors are present since the 
high-$p_\perp$ partons are close together in transverse space. They are 
colour rotated like a single parton. 

Under the further assumption that final state momenta of the high-$p_\perp$ 
partons are not resolved on the soft scale, the following result is derived,
\be
\frac{d\sigma}{dX^{(n+1)}}=\frac{k_0^2\,|T_H|^2}{\pi\,q_0\,N_c}\,
\int_{x_\perp}\,\left|\,\int_{k_\perp}\frac{\mbox{tr}[\tilde{W}_{x_\perp}
(k_\perp'\!\!\!-\!k_\perp)]}{k^2}\,\right|^2\delta^2\!\left(
\mbox{$\sum$}p_{(j)\perp}\right)\,\delta(q_0\!\!-\!q_0')\,.\label{cs2}
\ee
In this expression the non-perturbative input encoded in the Fourier 
transform $\tilde{W}$ has totally decoupled from the hard momenta that 
dominate $T_H$.

\section*{Results for scalar partons, quarks, and gluons}
The squared amplitude $|T_H|^2$ in Eq.~(\ref{cs2}) can be expressed through 
the partonic cross section $\hat{\sigma}(y)$ describing the collision of the 
photon, characterized by the standard variables $x$ and $Q^2$, with a parton 
carrying a fraction $y$ of the proton momentum. In Fig.~\ref{process} this 
corresponds to the interpretation of the line labelled by $k$ as an incoming 
line for the hard process. The cross section differential in $\xi=x(Q^2+M^2)
/Q^2$ takes the form
\be
\frac{d\sigma}{d\xi}=\int_x^\xi dy\,\hat{\sigma}(y)\left(\frac{df(y,\xi)}
{d\xi}\right)\,.
\ee
Introducing the variable $b\!=\!y/\xi$ the diffractive parton density for 
scalars reads 
\be
\left(\!\frac{df(y,\xi)}{d\xi}\!\right)_{\!\!scalar}\!\!\!\!\!=\frac{1}
{\xi^2}\left(\frac{b}
{1-b}\right)\int\frac{d^2k_\perp'(k_\perp'^2)^2}{(2\pi)^4N_c}\int_{x_\perp}
\left|\int\frac{d^2k_\perp}{(2\pi)^2}\,\frac{\mbox{tr}[\tilde{W}_{x_\perp}
(k_\perp'\!\!\!-\!k_\perp)]}{k_\perp'^2b+k_\perp^2(1-b)}\right|^2\,.
\ee

Analogous considerations with spinor and vector partons result in the same 
factorizing result, but with a diffractive quark density 
\be
\left(\frac{df(y,\xi)}{d\xi}\right)_{spinor}=\frac{2}{\xi^2}\int
\frac{d^2k_\perp'(k_\perp'^2)}{(2\pi)^4N_c}\int_{x_\perp}\left|\int
\frac{d^2k_\perp}{(2\pi)^2}\,\frac{k_\perp\mbox{tr}[\tilde{W}_{x_\perp}
(k_\perp'\!\!\!-\!k_\perp)]}{k_\perp'^2b+k_\perp^2(1-b)}\right|^2\, ,
\ee
or a diffractive gluon density
\begin{eqnarray}
\left(\frac{df(y,\xi)}{d\xi}\right)_{vector}\!\!\!&=&\frac{1}{\xi^2}
\left(\frac{b}{1-b}\right)\int\!\frac{d^2k_\perp'(k_\perp'^2)^2}{(2\pi)^4\,
(N_c^2-1)}\times
\\ \nonumber\\
&&\int_{x_\perp}\left|\int\frac{d^2k_\perp}{(2\pi)^2}\,\left(\delta^{ij}+
\frac{2k_\perp^ik_\perp^j}{k_\perp'^2}\left(\frac{1-b}{b}\right)\right)
\frac{\mbox{tr}[\tilde{W}_{x_\perp}^{\cal A}(k_\perp'\!\!\!-\!k_\perp)]}
{k_\perp'^2b+k_\perp^2(1-b)}\right|^2\,.\nonumber
\end{eqnarray}
Here $W^{\cal A}$ is defined like $W$ but with the $U$'s in the adjoint 
representation.

Note that the same kinematical picture arises if the soft interaction 
with the proton is modelled by two-gluon exchange \cite{nz}. While the 
kinematical factors associated with the different components of the photon 
wave function remain unchanged $\tilde{W}$ is substituted by a two-gluon 
form factor \cite{wust}.

\section*{Conclusions}
It has been shown that the semiclassical calculation in the proton rest 
frame is consistent with the concept of diffractive parton densities. The 
emerging physical picture is as follows. Leading twist diffraction arises 
from virtual photon fluctuations containing at least one low-$p_\perp$ 
parton. The interactions of the hard partons from the photon wave function 
with the proton colour field are kinematically irrelevant. In the Breit 
frame the elastic scattering of the low-$p_\perp$ parton off the field has 
to be reinterpreted as pair production by the fast proton. One parton from 
this colour singlet pair collides with the photon producing the above hard 
final state partons.

\end{document}